# DIAGONAL BASED FEATURE EXTRACTION FOR HANDWRITTEN ALPHABETS RECOGNITION SYSTEM USING NEURAL NETWORK

J.Pradeep[1], E.Srinivasan[2] and S.Himavathi[3]

[1,2]Department of ECE, Pondicherry College Engineering, Pondicherry, India.
`jayabala.pradeep@pec.edu, esrinivasan@pec.edu`
[3]Department of EEE, Pondicherry College Engineering, Pondicherry, India.
`himavathi@pec.edu`

*ABSTRACT*

*An off-line handwritten alphabetical character recognition system using multilayer feed forward neural network is described in the paper. A new method, called, diagonal based feature extraction is introduced for extracting the features of the handwritten alphabets. Fifty data sets, each containing 26 alphabets written by various people, are used for training the neural network and 570 different handwritten alphabetical characters are used for testing. The proposed recognition system performs quite well yielding higher levels of recognition accuracy compared to the systems employing the conventional horizontal and vertical methods of feature extraction. This system will be suitable for converting handwritten documents into structural text form and recognizing handwritten names.*

## 1. INTRODUCTION

Handwriting recognition has been one of the most fascinating and challenging research areas in field of image processing and pattern recognition in the recent years [1] [2]. It contributes immensely to the advancement of an automation process and can improve the interface between man and machine in numerous applications. Several research works have been focusing on new techniques and methods that would reduce the processing time while providing higher recognition accuracy [3].

In general, handwriting recognition is classified into two types as off-line and on-line handwriting recognition methods. In the off-line recognition, the writing is usually captured optically by a scanner and the completed writing is available as an image. But, in the on-line system the two dimensional coordinates of successive points are represented as a function of time and the order of strokes made by the writer are also available. The on-line methods have been shown to be superior to their off-line counterparts in recognizing handwritten characters due to the temporal information available with the former [4] [5]. However, in the off-line systems, the neural networks have been successfully used to yield comparably high recognition accuracy levels .Several applications including mail sorting, bank processing, document reading and postal address recognition require off-line handwriting recognition systems. As a result, the off-line handwriting recognition continues to be an active area for research towards exploring the newer techniques that would improve recognition accuracy [6] [7].

The first important step in any handwritten recognition system is pre-processing followed by segmentation and feature extraction. Pre-processing includes the steps that are required to shape the input image into a form suitable for segmentation [8]. In the segmentation, the input image is segmented into individual characters and then, each character is resized into m x n pixels towards the training network.

The Selection of appropriate feature extraction method is probably the single most important factor in achieving high recognition performance. Several methods of feature extraction for character recognition have been reported in the literature [9]. The widely used feature extraction methods are Template matching, Deformable templates, Unitary Image transforms, Graph description, Projection Histograms, Contour profiles, Zoning, Geometric moment invariants, Zernike Moments, Spline curve approximation, Fourier descriptors, Gradient feature and Gabor features.

An artificial neural Network as the backend is used for performing classification and recognition tasks. In the off-line recognition system, the neural networks have emerged as the fast and reliable tools for classification towards achieving high recognition accuracy [10]. Classification techniques have been applied to handwritten character recognition since the 1990s. These methods include statistical methods based on Bayes decision rule, Artificial Neural Networks (ANNs), Kernel Methods including Support Vector Machines (SVM) and multiple classifier combination [11], [12].

U. Pal *et al*, have proposed a modified quadratic classifier based scheme to recognize the off-line handwritten numerals of six popular Indian scripts [7].

Multilayer perceptron has been used for recognizing Handwritten English characters [13].The features are extracted from Boundary tracing and their Fourier Descriptors. Character is identified by analysing its shape and comparing its features that distinguish each character. Also an analysis has been carried out to determine the number of hidden layer nodes to achieve high performance of back propagation network. A recognition accuracy of 94% has been reported for handwritten English characters with less training time.

Dinesh et al [14] have used horizontal/vertical strokes, and end points as the potential features for recognition and reported a recognition accuracy of 90.50% for handwritten Kannada numerals. However, this method uses the thinning process which results in the loss of features.

U. Pal et al [15] have proposed zoning and directional chain code features and considered a feature vector of length 100 for handwritten numeral recognition and have reported a high level of recognition accuracy. However, the feature extraction process is complex and time consuming.

In this paper, a diagonal feature extraction scheme for the recognizing off-line handwritten characters is proposed. In the feature extraction process, resized individual character of size 90x 60 pixels is further divided into 54 equal zones, each of size 10x10 pixels. The features are extracted from the pixels of each zone by moving along their diagonals. This procedure is repeated for all the zones leading to extraction of 54 features for each character. These extracted features are used to train a feed forward back propagation neural network employed for performing classification and recognition tasks. Extensive simulation studies show that the recognition system using diagonal features provides good recognition accuracy while requiring less time for training.

The paper is organized as follows. In section II, the proposed recognition system is presented. The feature extraction procedure adopted in the system is detailed in the section III. Section IV describes the classification and recognition using feed forward back propagation neural network. Section V presents the experimental results and comparative analysis. In section VI, the proposed recognition system in Graphical User Interface is presented and finally, the paper is concluded in section VII.

International Journal of Computer Science & Information Technology (IJCSIT), Vol 3, No 1, Feb 2011

## 2. THE PROPOSED RECOGNITION SYSTEM

In this section, the proposed recognition system is described. A typical handwriting recognition system consists of pre-processing, segmentation, feature extraction, classification and recognition, and post processing stages. The schematic diagram of the proposed recognition system is shown in Fig.1

### 2.1. Image Acquisition

In Image acquisition, the recognition system acquires a scanned image as an input image. The image should have a specific format such as JPEG, BMT etc.  This image is acquired through a scanner, digital camera or any other suitable digital input device.

### 2.2. Pre-processing

The pre-processing is a series of operations performed on the scanned input image. It essentially enhances the image rendering it suitable for segmentation. The various tasks performed on the image in pre-processing stage are shown in Fig.2. Binarization process converts a gray scale image into a binary image using global thresholding technique. Detection of edges in the binarized image using sobel technique, dilation the image and filling the holes present in it are the operations performed in the last two stages to produce the pre-processed image suitable for segmentation [16].

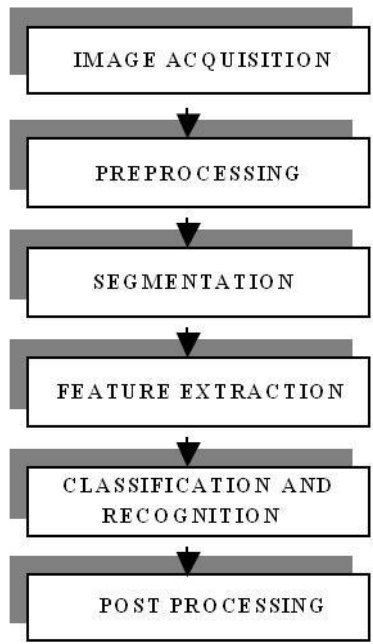

Figure 1.  Schematic diagram of the proposed recognition system

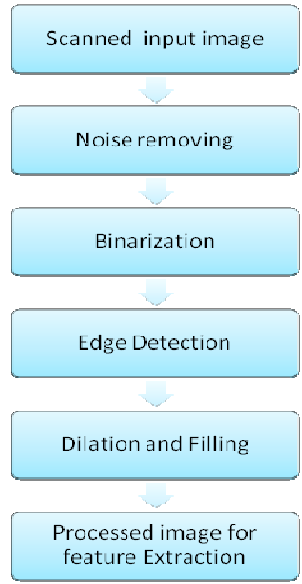

Figure 2.  Pre-processing of handwritten character

### 2.3. Segmentation

In the segmentation stage, an image of sequence of characters is decomposed into sub-images of individual character. In the proposed system, the pre-processed input image is segmented into isolated characters by assigning a number to each character using a labelling process. This labelling provides information about number of characters in the image. Each individual character is uniformly resized into 90X60 pixels for classification and recognition stage.

## 3. PROPOSED FEATURE EXTRACTION METHOD

In this stage, the features of the characters that are crucial for classifying them at recognition stage are extracted. This is an important stage as its effective functioning improves the recognition rate and reduces the misclassification [17]. Diagonal feature extraction scheme for recognizing off-line handwritten characters is proposed in this work. Every character image of size 90x 60 pixels is divided into 54 equal zones, each of size 10x10 pixels (Fig.3(c)). The features are extracted from each zone pixels by moving along the diagonals of its respective 10X10 pixels. Each zone has19 diagonal lines and the foreground pixels present long each diagonal line is summed to get a single sub-feature, thus 19 sub-features are obtained from the each zone. These 19 sub-features values are averaged to form a single feature value and placed in the corresponding zone (Fig.3 (b)). This procedure is sequentially repeated for the all the zones. There could be some zones whose diagonals are empty of foreground pixels. The feature values corresponding to these zones are zero. Finally, 54 features are extracted for each character. In addition, 9 and 6 features are obtained by averaging the values placed in zones rowwise and columnwise, respectively. As result, every character is represented by 69, that is, 54 +15 features.

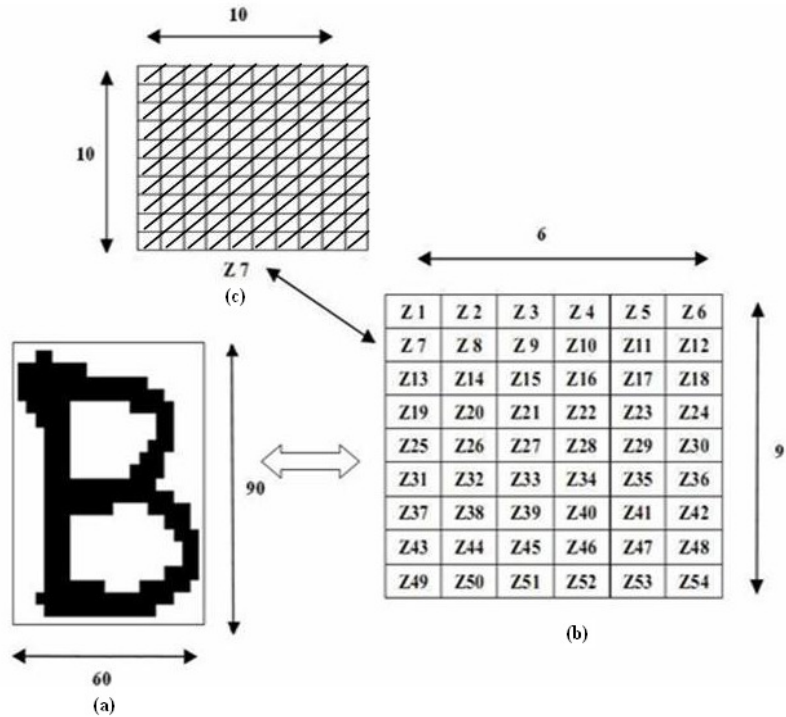

Figure 3. Procedure for extracting feature from the characters

## 4. CLASSIFICATION AND RECOGNITION

The classification stage is the decision making part of a recognition system and it uses the features extracted in the previous stage. A feed forward back propagation neural network having two hidden layers with architecture of 54-100-100-38 is used to perform the classification. The hidden layers use log sigmoid activation function, and the output layer is a competitive layer, as one of the characters is to be identified. The feature vector is denoted as X where $X = (f_1, f_2,\ldots,f_d)$ where $f$ denotes features and $d$ is the number of zones into which each character is divided. The number of input neurons is determined by length of the feature vector d. The total numbers of characters n determines the number of neurons in the output layer. The number of neurons in the hidden layers is obtained by trial and error. The most compact network is chosen and presented.

The network training parameters are:

- Input nodes : 54/69
- Hidden nodes : 100 each
- Output nodes :38  (26 alphabets, 10 numerals and 2 special symbols)
- Training algorithm : Gradient descent with momentum training and adaptive learning
- Perform function :Mean Square Error
- Training goal achieved : 0.000001
- Training epochs :1000000
- Training momentum constant: 0.9.

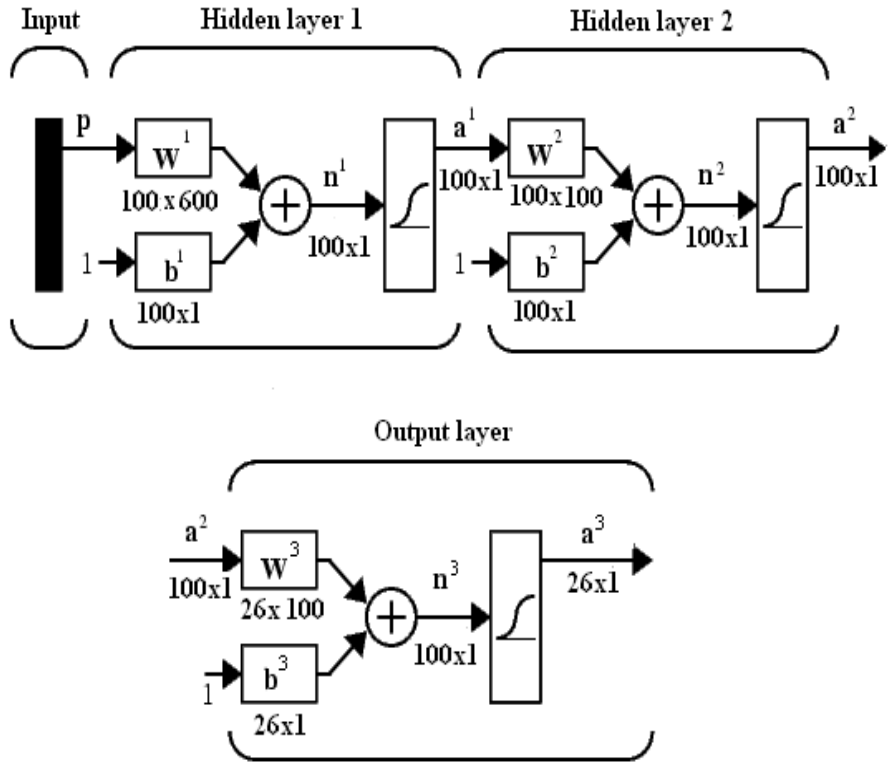

Figure 4. Three layer neural network for character recognition

The architecture of the network with two layers is illustrated in fig. 4. The output of $i^{th}$ layer is given by

$$a^i = \log sig(w^i a^{i-1} + b^i) \tag{1}$$

where,

$i = [1, 2, 3]$ and $a^0 = P$

$w^i =$ Weight vector of $i^{th}$ layer

$a^i =$ Output of $i^{th}$ layer

$b^i =$ Bias vector for $i^{th}$ layer

## 5. RESULTS AND DISCUSSION

The recognition system has been implemented using Matlab7.1.The scanned image is taken as dataset/ input and feed forward architecture is used. The structure of neural network includes an input layer with 54/69 inputs, two hidden layers each with 100 neurons and an output layer with 26 neurons. The gradient descent back propagation method with momentum and adaptive learning rate and log-sigmoid transfer functions is used for neural network training. Neural network has been trained using known dataset. A recognition system using two different feature lengths is built. The number of input nodes is chosen based on the number of features. After training the network, the recognition system was tested using several unknown dataset and the results obtained are presented in this section.

Two approaches with three different ways of feature extraction are used for character recognition in the proposed system. The three different ways of feature extraction are horizontal direction, vertical direction and diagonal direction**.**

In the first approach, the feature vector size is chosen as 54, i.e. without rowwise and columnwise features. The results obtained using three different types of feature extraction are summarized in Table. I. The criteria for choosing the type of feature extraction are: (i) the speed of convergence, i.e. number of epochs required to achieve the training goal and (ii) training stability. However, the most important parameter of interest is the accuracy of the recognition system. The results presented in Table 1 show that the diagonal feature extraction yields good recognition accuracy compared to the others types of feature extraction. Fig.5.shows the Error (MSE) vs. Training Epochs performance function of the network with 54 features obtained though diagonal extraction. The desired performance goal has been achieved in 923 epochs.

In the second approach, the feature vector size is chosen as 69 including the rowwise and columnwise features. The results obtained for the second approach is also presented in Table. II and it shows that the diagonal feature extraction provides higher recognition accuracy, compared to the others types of feature extraction. Fig.6. shows the Error (MSE) vs. Training Epochs performance of the network with 69 features obtained though diagonal extraction. It can be noted that it requires 854 epochs to reduce the mean square error to the desired level.

TABLE.I
COMPARISON OF RECOGNITION RATE RESULTS OBTAINED USING DIFFERENT ORIENTATIONS WITH 54 FEATURES

| Networks | 1 | 2 | 3 |
|---|---|---|---|
| **Feature Extraction type** | **Vertical** | **Horizontal** | **Diagonal** |
| **Number of nodes in input layer** | 54 | 54 | 54 |
| **Number of nodes in 1st hidden layer** | 100 | 100 | 100 |
| **Number of nodes in 2st hidden layer** | 100 | 100 | 100 |
| **Number of nodes in output layer** | 26 | 26 | 26 |
| **Recognition rate percentage** | 92.69 | 93.68 | 97.80 |

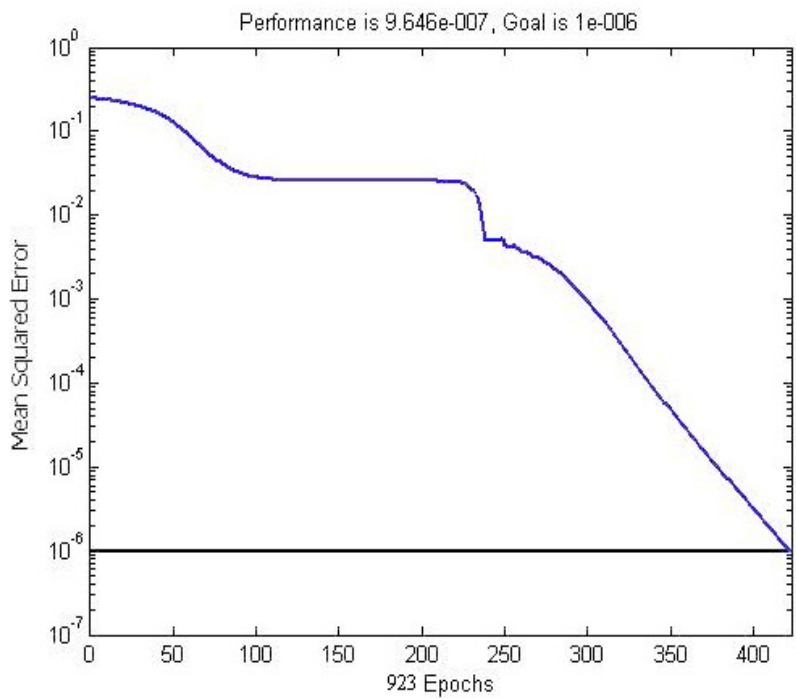

Figure 5. The variation of MSE with training Epochs for 54 features

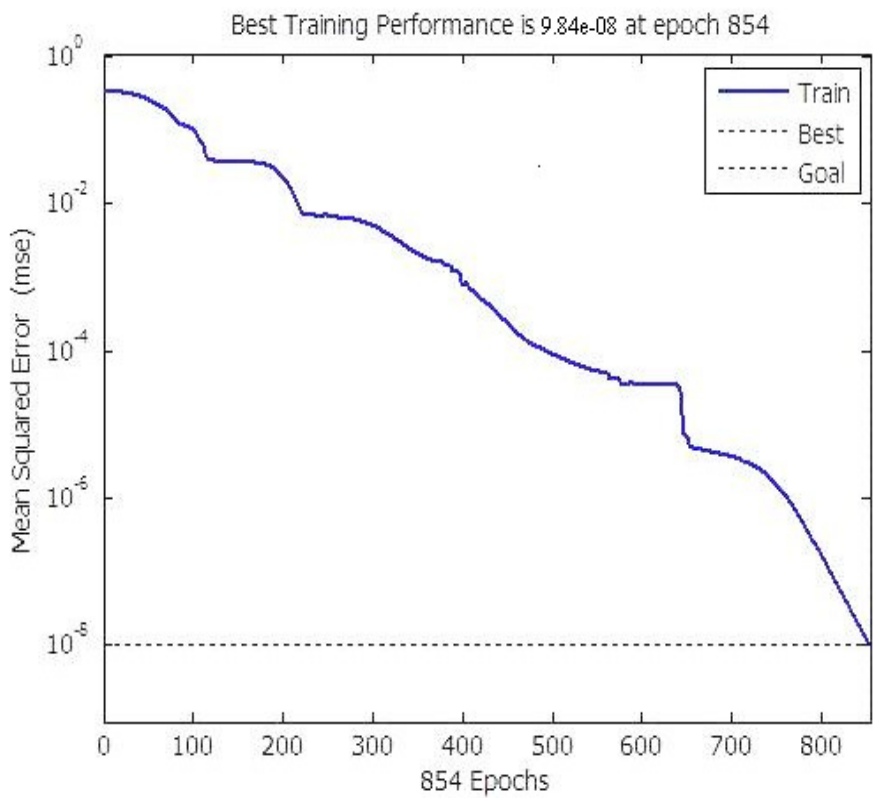

Figure 6. The variation of MSE with training Epochs for 69 features

TABLE.II
COMPARISON OF RECOGNITION RATE RESULTS OBTAINED USING
DIFFERENT ORIENTATIONS WITH 69 FEATURES

| Networks | 1 | 2 | 3 |
|---|---|---|---|
| Feature Extraction type | Vertical | Horizontal | Diagonal |
| Number of nodes in input layer | 69 | 69 | 59 |
| Number of nodes in 1st hidden layer | 100 | 100 | 100 |
| Number of nodes in 2st hidden layer | 100 | 100 | 100 |
| Number of nodes in output layer | 26 | 26 | 26 |
| Recognition rate percentage | 92.69 | 94.73 | 98.54 |

## 6. IMPLEMENTATION ON GRAPHICAL USER INTERFACE

A user-friendly front end interface as shown in Fig.7 and Fig.8 has been implemented for the proposed handwritten character recognition system using menu based GUI (Graphical User Interface). The interface system presents the user with two menus - first menu with five processing stages (Fig.7) and the second menu to choose the type feature extraction (Fig.8).

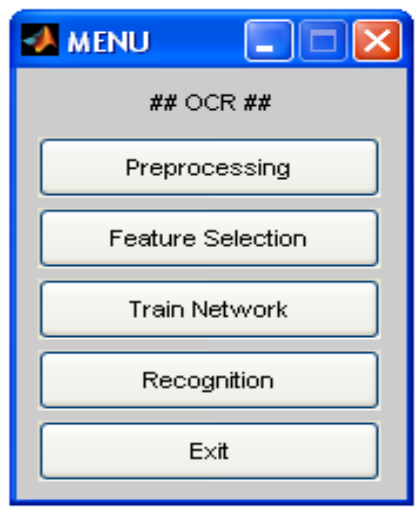

Figure 7. Recognition System using Menu Based Graphical User Interface

The menu based GUI enables the user to perform pre-processing, select the type of feature extraction, perform the feature extraction using the chosen method and train the network. After the network is trained, the recognition of the test image can be initiated by clicking the recognition bar on the interface. The test image is chosen using the facility provided for selecting the test images. Upon the completion of recognition process, the recognized image appears on the notepad. The same procedure can be used to recognize any number of test images. Finally the exit bar is used to quit from the character recognition system after recognizing all the test images. The GUI frees the user from the difficulties of working from the command line interface.

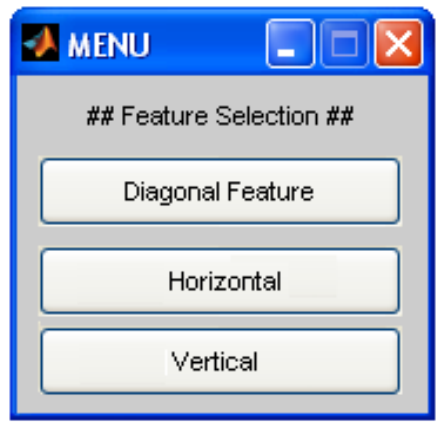

Figure 8. Feature Extraction Options in Menu Based Graphical User Interface

## 7. CONCLUSION

A simple off-line handwritten English alphabet characters recognition system using a new type of feature extraction, namely, diagonal feature extraction is proposed. Two approaches using 54 features and 69 features are chosen to build the Neural Network recognition system. To compare the recognition efficiency of the proposed diagonal method of feature extraction, the neural network recognition system is trained using the horizontal and vertical feature extraction methods. Six different recognition networks are built. Experimental results reveals that 69 features gives better recognition accuracy than 54 features for all the types of feature extraction. From the test results it is identified that the diagonal method of feature extraction yields the highest recognition accuracy of 97.8 % for 54 features and 98.5% for 69 features. The diagonal method of feature extraction is verified using a number of test images. The proposed off-line hand written character recognition system with better-quality recognition rates will be eminently suitable for serval applications including postal/parcel address recognition, bank proecssing, document reading and conversion of any handwritten document into structural text form.

International Journal of Computer Science & Information Technology (IJCSIT), Vol 3, No 1, Feb 2011International Journal of Computer Science & Information Technology (IJCSIT), Vol 3, No 1, Feb 2011

*Journal of Theoretical and Applied Information Technology*, JATIT vol.4, no.12, pp.1171-1181, 2008.

**Authors**

**J.Pradeep** received his B.Tech degree in Electronics and Communication Engineering from Barathiyar college of Engineering and Technogy affiliated to Pondicherry University in the year 2005. He obtained his M.Tech degree in Electronics and Communication Engineering from Podicherry Engineering College in the year 2009. He is currently a Ph.D candidate in the Department of Electronics and Communication Engineering in Podicherry Engineering College. He has published two papers in International Journal. He is a life member of ISTE. His areas of interest are Wireless Communication, Image proceesing and Neural networks.

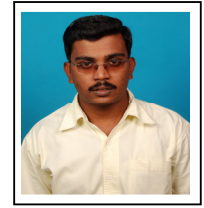

**E.Srinivasan** obtained his B.E. degree in Electrical and Electronics Engineering from P.S.G. College of Technology, Coimbatore, India, in the year 1984. He received his M.E. degree in Instrumentation Technology in the year 1987 from Madras Institute of Technology, Chennai, India. He was awarded with Ph.D. degree by the Anna University, Chennai, India in the year 2003 for his research contributions in Nonlinear Signal Processing. Currently, he is serving as Professor and Head of the Department of Electronics and Communication Engineering, Pondicherry Engineering College, Pondicherry, India. He has published 30 research papers in national/international journals/conferences. He is a reviewer of the AMSE journal of Signal Processing. His research interests include nonlinear signal processing and pattern recognition and their applications.

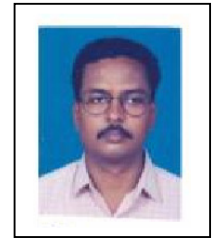

**S.Himavathi** completed her BE degree in Electrical and Electronics Engineering from College of Engineering, Guindy, Chennai, India, in the year 1984. She obtained her M.E. degree in Instrumentation Technology in the year 1987 from Madras Institute of Technology, Chennai, India. She completed her Ph.D. degree in the area of Fuzzy modeling in the year 2003 from Anna University, Chennai, India. She is a Professor and Head of the Department of Electrical and Electronics Engineering, Pondicherry Engineering College. She has around 50 publications to her credit. She is a reviewer of the AMSE journal of Modeling, IEEE Industrial Electronics Society and Asian Neural Networks Society. Her research interests are Fuzzy systems, Neural Networks, Hybrid systems and their applications.

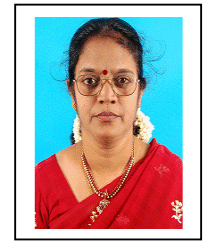